\documentstyle[mprocl]{article} 

\input{psfig.sty}

\newcommand{\multcc}[1]{\multicolumn{1}{c}{#1}}
\def\lsim{\hbox{ \rlap{\raise 0.425ex\hbox{$<$}}\lower 0.65ex\hbox{$\sim$} }}
\def\gsim{\hbox{ \rlap{\raise 0.425ex\hbox{$>$}}\lower 0.65ex\hbox{$\sim$} }}
\def\etal{{\it et al. }}

\def\be{\begin{equation}}
\def\ee{\end{equation}}
\def\bea{\begin{eqnarray}}
\def\eea{\end{eqnarray}}

\begin{document}

\title{THE COSMIC GAMMA-RAY BURSTS
\footnote{Based in part on the observations obtained at the W.M.~Keck 
Observatory, operated by the California Association for Research in Astronomy,
a scientific partnership among Caltech, the Univ. of California and NASA;
and with the NASA/ESA Hubble Space Telescope, operated by the AURA, Inc., 
under a contract with NASA.
}
\footnote{\bf To appear in Proc. IX Marcel Grossmann Meeting,
eds. Gurzadyan, V., Jantzen, R., and Ruffini, R.,
Singapore: World Scientific, in press (2001).}
}

\author{
S. G. DJORGOVSKI $^\star$,
D. A. FRAIL $^\dag$,
S. R. KULKARNI $^\star$,
R. SARI $^\ast$,
}
\author{
J. S. BLOOM $^\star$,
T. J. GALAMA $^\star$,
F. A. HARRISON $^\ddag$,
P. A. PRICE $^\ddag$,
}
\author{
D. FOX $^\star$,
D. E. REICHART $^\star$,
S. YOST $^\ddag$,
E. BERGER $^\star$,
}
\author{
A. DIERCKS $^\star$,
R. GOODRICH $^\S$,
\& F. CHAFFEE $^\S$
}

\address{}

\address{$^\star$ Palomar Observatory, Caltech, Pasadena, CA 91125, USA \\
         $^\dag$ NRAO/VLA, Socorro, NM 87801, USA \\
         $^\ast$ Theoretical Astrophysics, Caltech, Pasadena, CA 91125, USA \\
         $^\ddag$ Space Radiation Laboratory, Caltech, Pasadena, CA 91125, USA \\
         $^\S$ W.M.~Keck Observatory, CARA, Kamuela, HI 96743, USA \\
}

\maketitle
\abstracts{ 
Cosmic $\gamma$-ray bursts are one of the great frontiers of astrophysics today.
They are a playground of relativists and observers alike.  They may teach us
about the death of stars and the birth of black holes, the physics in
extreme conditions, and help us probe star formation in the distant and
obscured universe.  In this review we summarise some of the remarkable progress
in this field over the past few years.  
While the nature of the GRB progenitors is still unsettled, it now appears
likely that at least some bursts originate in explosions of very massive stars,
or at least occur in or near the regions of massive star formation.  
The physics of the burst afterglows is reasonably well understood, and has 
been tested and confirmed very well by the observations.  
Bursts are found to be beamed, but with a broad range of jet opening angles;
the mean $\gamma$-ray energies after the beaming corrections are $\sim 10^{51}$
erg. 
Bursts are associated with faint ($\langle R \rangle \sim 25$ mag) galaxies at
cosmological redshifts, with $\langle z \rangle \sim 1$. 
The host galaxies span a range of luminosities and morphologies,
but appear to be broadly typical for the normal, actively star-forming galaxy
populations at comparable redshifts and magnitudes. 
Some of the challenges for the future include: the nature of the short bursts
and possibly other types of bursts and transients; use of GRBs to probe the
obscured star formation in the universe, and possibly as probes of the very
early universe; and their detection as sources of high-energy particles and
gravitational waves. 
}

\section{Introduction}

Ever since their discovery \cite{kle73}, the nature of the cosmic
$\gamma$-ray bursts (GRBs) has been one of the great puzzles of science.  
While a complete physical explanation of this remarkable phenomenon is still
not in hand, there has been a great deal of progress in this field over the
last few years.  Studies of GRBs are now one of the most active areas of
research in astronomy, with a publication rate of $\sim 500$ per year
\cite{hur00}, and the total number of GRB-related publications now exceeding
5000.  GRBs represent a great laboratory in the sky for relativistic
astrophysics, with Lorentz factors reaching $\Gamma \sim 10^2 - 10^3$. 

The pre-1997 state of affairs was summarised well in many reviews
\cite{fis95,lam95,pac95}.  
The distribution of bursts on the sky found by BATSE/CGRO is highly isotropic
\cite{mee92}, which provided the first solid hint about their cosmological
origins. 
After many years of speculation based on a limited observational evidence,
handicapped mainly by the lack of precise and rapid positional identifications,
the field was revolutionized by the BeppoSAX satellite \cite{boe97}. 
The key was the enabling discovery of long-lived and precisely located GRB
afterglows at longer wavelengths, in the X-rays \cite{cfh+97}, 
optical \cite{vgg+97}, and radio \cite{fkn+97}, and the resulting direct
determination of the cosmological distance scale to the bursts \cite{mdk+97}. 

Allowing for the observational selection and coverage, GRBs are detectable 
at a mean rate of $\sim 10^3$ per year down to the limiting 
fluxes of $\sim 10^{-7}$ erg cm$^{-2}$ s$^{-1}$, 
or fluences of $\sim 10^{-6}$ erg cm$^{-2}$.  
While there must be many fainter bursts, the flattening of the observed flux
distribution suggests that their numbers are at most comparable
\footnote{ We note that the nature of GRB detection data (time series)
precludes detection of low amplitude, long duration bursts, if they do exist;
this is a time-domain analogue of the familiar image-domain low surface
brightness bias.}.
The cumulative distribution of fluxes indicates a radially non-uniform
(i.e., non-Euclidean or evolving) spatial distribution \cite{kom00}.

Since we now know that the observed bursts are (at least in principle)
detectable out to proper distances in excess of $10^{28}$ cm, or redshifts $z
\sim 10$, i.e., reaching the earliest phases of galaxy formation, in rough
numbers the bursts occur at a rate of a few per universe per day, 
or once per few million years per average galaxy, 
or $\sim 10^{-91}$ cm$^{-3}$ s$^{-1}$ ...
These numbers do not include any beaming corrections (see below), which would
increase the actual rates (at the expense of the implied energetics per burst)
by a factor of a few hundred.
Comparison of the observed GRB rate with those of other astrophysical 
phenomena, e.g., supernov\ae, which occur on average once per century per
typical galaxy
\footnote{ Note, however, that the SN rate from very massive progenitors must
be much lower, and can be comparable to the GRB rate.},
or the expected rate of neutron star mergers, which may occur at a rate of
$\sim 10^{-6}$ per year per typical galaxy,
etc.,
can be used to constrain the models of their origin.

While the bursts are the brightest sources on the $\gamma$-ray sky when they
do occur (and could, at least briefly, achieve luminosities comparable to
all of the rest of the universe, $\sim 10^{54}$ erg s$^{-1}$), they do not
last very long, and are not known to be a major contributor to the overall
energy production in the universe: the observed local energy density due to
GRBs (independent of the beaming corrections) is 
a few $\times 10^{-21}$ erg cm$^{-3}$, 
i.e., some 4 orders of magnitude less than the cosmic X-ray background, 
6 orders of magnitude less than that of all the starlight ever emitted, 
and 6 orders of magnitude less than the local energy density of the CMBR. 

Their high-energy spectra are featureless continua, typically a broken
power-law, suggesting a non-thermal (e.g., synchrotron) origin, with the peak
energies near $\sim 0.1 - 1$ MeV.  An interesting and as yet unsettled question
is whether other, similar populations of transients may exist at lower energies
(e.g., X-ray bursts).  It is also not yet known what are the maximum energies
of particles generated in bursts, but the TeV range certainly seems viable. 

An important hint about their possible origins is provided by the timescales of
the bursts ($\sim 0.1 - 100$ s), and the intrinsic variability ($\sim 10^{-3}$
s), suggesting a spatial scale comparable to that of stars and dense stellar
remnants, i.e., black holes (BH) or neutron stars (NS).  This is probably the
strongest argument in favor of stellar origin models for GRBs. 

Since distances to the bursts have now been measured for many cases (see
below), the implied isotropic $\gamma$-ray energies span the range
$E_{\gamma,iso} \sim 10^{51} - 10^{54}$ erg $\approx 10^{-3} - 1 ~M_\odot c^2$.
At least at the lower end of the range, this is comparable to the energy
release in supernov\ae, which again suggests (but does not compel) a
physical connection with the death of massive stars, or at least the birth
of stellar mass black holes.
The implied mean energy density at the peak can reach that at the time of 
$\sim 10^{-3}$ s after the big bang, i.e., at the onset of the cosmic
baryogenesis.

There are several distinct physical stages of a GRB, each with its own range
of scales.  The ratio of the empirically based understanding and knowledge to
speculation $\rightarrow 0$ as $(t-t_{burst}) \rightarrow 0$.
\begin{enumerate}
\item {\bf The Prime Mover:}~ $t \sim 10^{-3}$ s (?).  While plausible models 
      do exist, this is fundamentally unknown.  It is possible that future 
      observations of gravitational waves or high-energy cosmic rays and 
      neutrinos would bring some direct insight into the central engine of GRBs.
\item {\bf The Burst Itself:}~ $t \sim 10^{-1} - 10^2$ s.  This is mainly 
      probed by the burst $\gamma$-ray light curves and spectra.  Some physical
      understanding of this phase (internal shocks, etc.) exists.
\item {\bf The Afterglow:}~ $t \sim 10$ s $\rightarrow \infty$.  This is the
      best probed and best understood phase, with detailed observations over
      a broad range of wavelengths and time scales.  In many ways, the
      physics and the behavior of afterglows are independent of the actual
      physical mechanism behind the bursts: given a certain (large!) amount 
      of energy deposited in a relatively small volume, a relativistic
      fireball is inevitable, with subsequent expansion, shocks, etc.
\end{enumerate}
In addition, observations of GRB host galaxies (or their environments in
general), possible association with coincident supernov\ae, etc., can provide
indirect clues about the nature of progenitors. 

Well over a 100 distinct theoretical models for GRBs have been proposed, with
many more sub-variants \cite{nem94}.  
The establishment of the cosmological distance scale to the GRBs (and therefore
the energetics scale) has focused the debate considerably.  While many
interesting, novel (and possibly even correct) ideas continue to be generated 
(see, e.g., \cite{ruf2000}), the prevailing view nowadays is focused on two
types of models: 
explosions of very massive ($> 30~M_\odot$?) stars (also known as ``collapsar''
or ``hypernova'' type models \cite{pac98b,mfw99}), 
and mergers of compact stellar remnants \cite{eic+89}
(NS, BH, or even white dwarfs; but with at least one mergee being a NS or a
BH). 

In both cases, the end product is a stellar mass scale BH, surrounded by a
rapidly rotating torus, whose orbital kinetic energy can be extracted 
via MHD processes and used to power the GRB.  If the BH itself is threaded
by the magnetic field (which has to be amplified to $\sim 10^{15}$ G!),
its spin energy can be extracted via the Blandford-Znajek mechanism
\cite{BZ77}.  Both mechanisms can extract $\sim 10^{54}$ erg, and both
provide a natural collimation (spin) axis, for energy release via Poynting
jets.  Additional energy ($\sim 10^{51}$ erg) can be provided by thermal
neutrino cooling, $\nu \bar\nu \rightarrow e^\pm, \gamma$. 
The gravitational wave component is strongly model-dependent and is highly
uncertain at this point; hopefully it will be settled observationally with
LIGO \footnote{ http://www.ligo.caltech.edu/}
and LISA \footnote{ http://lisa.jpl.nasa.gov/}.

Regardless of the exact model for GRBs, it appears highly likely that black
holes are involved.  While quasars (and AGN in general) represent probes of
the BH physics in the $M_\bullet \sim 10^{6-9} ~M_\odot$ range, 
GRBs can be a powerful observational probe of the BH physics on the stellar
mass scales, $M_\bullet \sim 10^{0.5 - 1.5} ~M_\odot$, and on correspondingly
shorter time scales.

As we will see bellow, the evidence for the collapsar/hypernova type of models
is becoming increasingly compelling, at least for the well-studied long
bursts, but the case is still not closed.  It is entirely possible that more
than one physical model is at work, by analogy with nov\ae\ and supernov\ae,
where very different physical mechanisms lead to a roughly comparable observed
phenomenology.  On that agnostic note, we will not summarise or even reference
any particular models (thus offending all of the relevant authors equally), 
and simply direct the interested reader to recent
reviews  \cite{mesz99,mesz00,mesz01,piran99,piran01}. 

Much of the pre-1997 work was based on the 
CGRO mission \footnote{ http://cossc.gsfc.nasa.gov/}
and the IPN network \footnote{ http://ssl.berkeley.edu/ipn3/}.
Today, several space missions and a large number of ground-based telescopes
and teams are studying bursts and their afterglows
\footnote{ We note that all of the recent progress has been in the follow-up
of the so-called long bursts, with 90-th percentile durations measured in
seconds to tens of seconds.  We do not really know much yet about the
short bursts family, with typical durations $\sim 0.1$ s.}.
Another critically important technological development in studies of GRBs is the
rapid dissemination of time-critical observations via email and WWW, mainly
through the {\it GCN Circulars} (S. Barthelmy, NASA GSFC)
\footnote{ http://gcn.gsfc.nasa.gov/gcn/gcn3\_archive.html}.
An excellent on-line archive of GRB observations is maintained by J. Greiner
(Astr. Inst. Potsdam)
\footnote{ http://www.aip.de/People/JGreiner/grbgen.html}.

In this review we summarise some of the recent developments, mainly from an
observational point of view, as of the early/mid-2001.  We apologize for any
undue omissions and incompleteness.  Other recent reviews \cite{vkw00,kbb+00}
provide additional information.

\section{GRB Afterglows: Physics and Observations}

The extreme characteristics of GRBs lead to a paradox (the
``compactness problem'').  Assuming that $\sim 10^{52}$ erg worth of
photons, distributed according to the GRB spectrum, is released in a
small volume of linear dimensions $R\le c \Delta t$, then the optical depth for
pair creation is $\tau \sim 10^{15}$.  If so, all the photons would
have interacted to create pairs and thermalize.  However, the observed spectrum
of GRBs is highly non-thermal!  The only known solution to this problem is
relativistic motion.  If the emission site is moving relativistically, with a
Lorentz factor $\Gamma$, toward the observer, then the optical depth is
reduced, compared to the stationary estimate, due to two effects: First, the
size of the source can be bigger by a factor of $\Gamma^{2}$. This will still
produce variability over a short time scale given by $\Delta t = R/\Gamma^{2}c$
since not all the source is seen as the radiation for a relativistically moving
object is beamed.  Second, the photons in the local frame are softer by a
factor of $\Gamma$, and therefore only a small fraction of them, at the high
energy tail, have enough energy to create pairs.  The combination of these two
effects reduces the optical depth by a factor of $\sim \Gamma ^{6.5}$. 
Therefore, the optical depth is reduced below unity if $\Gamma > 100$ or so. 

This leads to a generic scenario for GRBs.  First, a compact source (``the
prime mover'') releases $\sim 10^{52}$ erg in a small volume of space ($\sim
10^7$ cm?) and on a short time scale ($\sim 0.1 - 100$ s).  This large
concentration of energy (a relativistic fireball) expands due to its own
pressure; particle pairs are produced and coupled to the radiation field.  
If the rest mass present within the burst region is not too large, $\leq
10^{-5}M_{\odot }$, this will result in relativistic expansion with
$\Gamma>100$ (the requirement for such a peculiar, small baryonic contamination
presents an interesting fine-tuning problem by itself).  Finally, at a large
enough radius, the kinetic energy of the expanding material is converted to
internal energy and radiated, mainly in $\gamma$-rays.  At this stage the system
is optically thin and high energy photons can escape. 

In order to convert the kinetic energy into photons, two scenarios were
proposed: external shocks \cite{MR93} and internal shocks \cite{NPP92,RM94}. 
In the external shocks scenario, the relativistic material is running into some
(external) ambient medium, probably the interstellar medium or a wind that was
emitted earlier by the progenitor.  In the internal shocks scenario the inner
engine is assumed to emit an irregular flow, that consists of many shells, that
travel with a variety of Lorentz factors and therefore colliding into each
other and thermalizing some of their kinetic energy. 
Observed variability in most GRBs provides a way of distinguishing between
the two scenarios.  External shocks
require a complicated surrounding with a relatively simple source that
explodes once, while internal shocks require a more complicated source
that will explode many times to produce several shells.
A variety of arguments \cite{FMN96,SP97} favors the internal shock model.

We also note that in the currently favorite models for the central engine (with
the production of a rapidly rotating BH and a torus), the orbital time scales
($\sim 10^{-3}$ s) are several orders of magnitude shorter than the observed
burst time scales, thus imposing an interesting stability problem. 

While the detailed understanding of the physics of the bursts is still not
complete, and the nature of the triggers is still largely hypothetical,
we do have a reasonable physical understanding of the subsequent stages
of the phenomenon, i.e., the burst afterglows.  The afterglows were predicted
well before they were observed \cite{PR93,K94,V97,MR97}.
The afterglow theory is relatively simple, and it has been confirmed
(at least in a broad sense)
spectacularly well by the observations.  It deals with the emission on much
longer timescales, and thus the poorly known details and complexities of the
initial conditions are relatively unimportant,
and the physical description of afterglows depends on a small number of
parameters, such as the total energy and the density of the external medium.

After the internal shocks produced the GRB itself, the expanding shells
interact with the surrounding medium and decelerate.  The emission shifts into
lower and lower frequencies. 
The observed afterglows usually show a power-law decay $t^{-\alpha }$ in the
optical and X-ray where a typical value is $\alpha \approx 1.2$.  Some
afterglows show a steeper decline with $\alpha\approx-2$.  In the radio
wavelengths, the flux seems to rise on timescale of weeks and then decays with
a similar power-law. 

\begin{figure}
\centerline{\psfig{figure=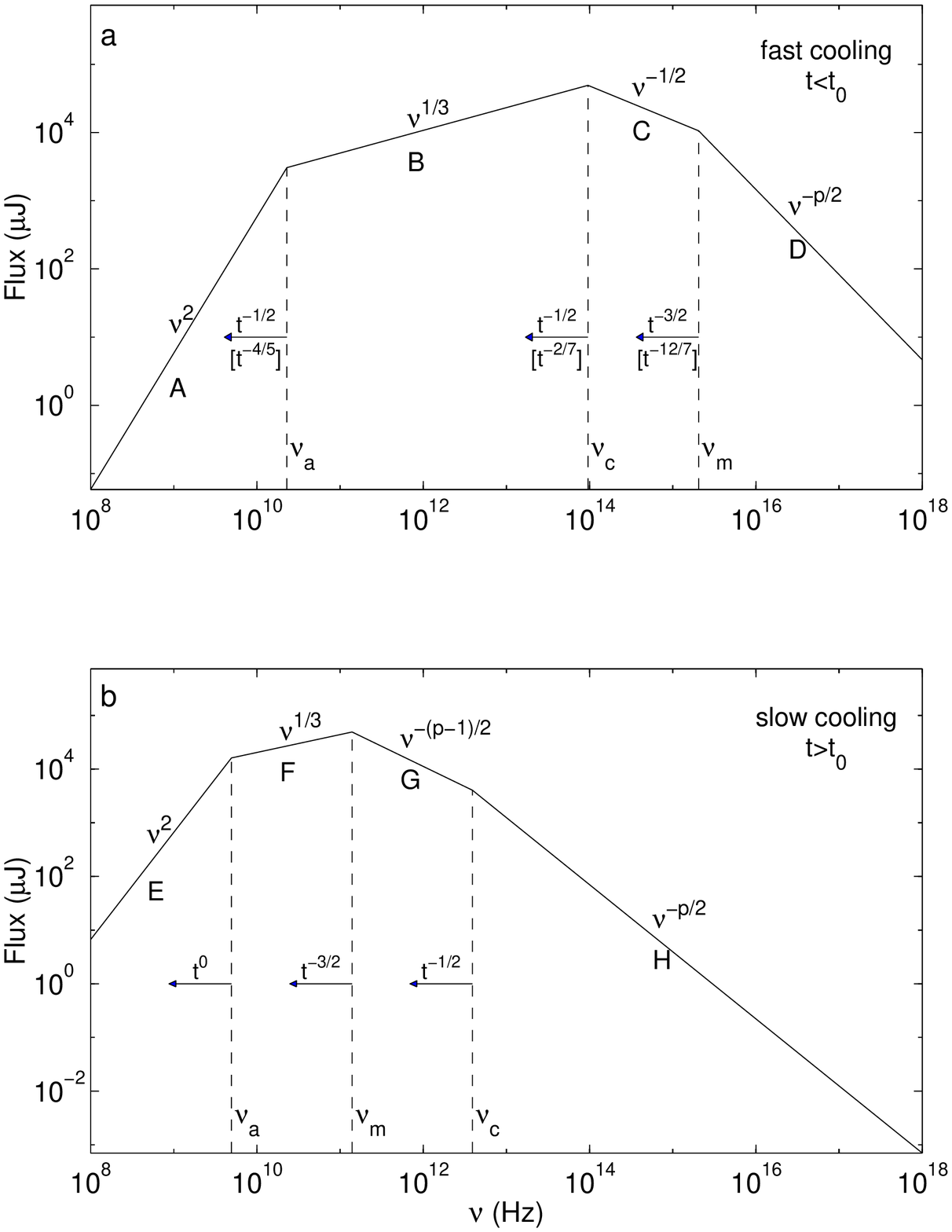,height=5.0in}}
\caption[]{
Theoretical spectra of synchrotron emission from a power-law distribution of
electrons.  
(a) Fast cooling, which is expected at early times.  The characteristic
frequencies decrease with time as indicated; the scalings above the arrows
correspond to an adiabatic evolution, and the scalings below, in square
brackets, correspond to a fully radiative evolution. 
(b) Slow cooling, which is expected at late times.  The evolution is always
adiabatic.  
Electron energy power-law index $p \approx 2.2-2.4$ fits well the observed
spectra.  The temporal scalings correspond to the case of a spherical
fireball shock expanding into a constant density medium.
From \cite{SPN98}. 
}
\end{figure}

\begin{figure}
\centerline{\psfig{figure=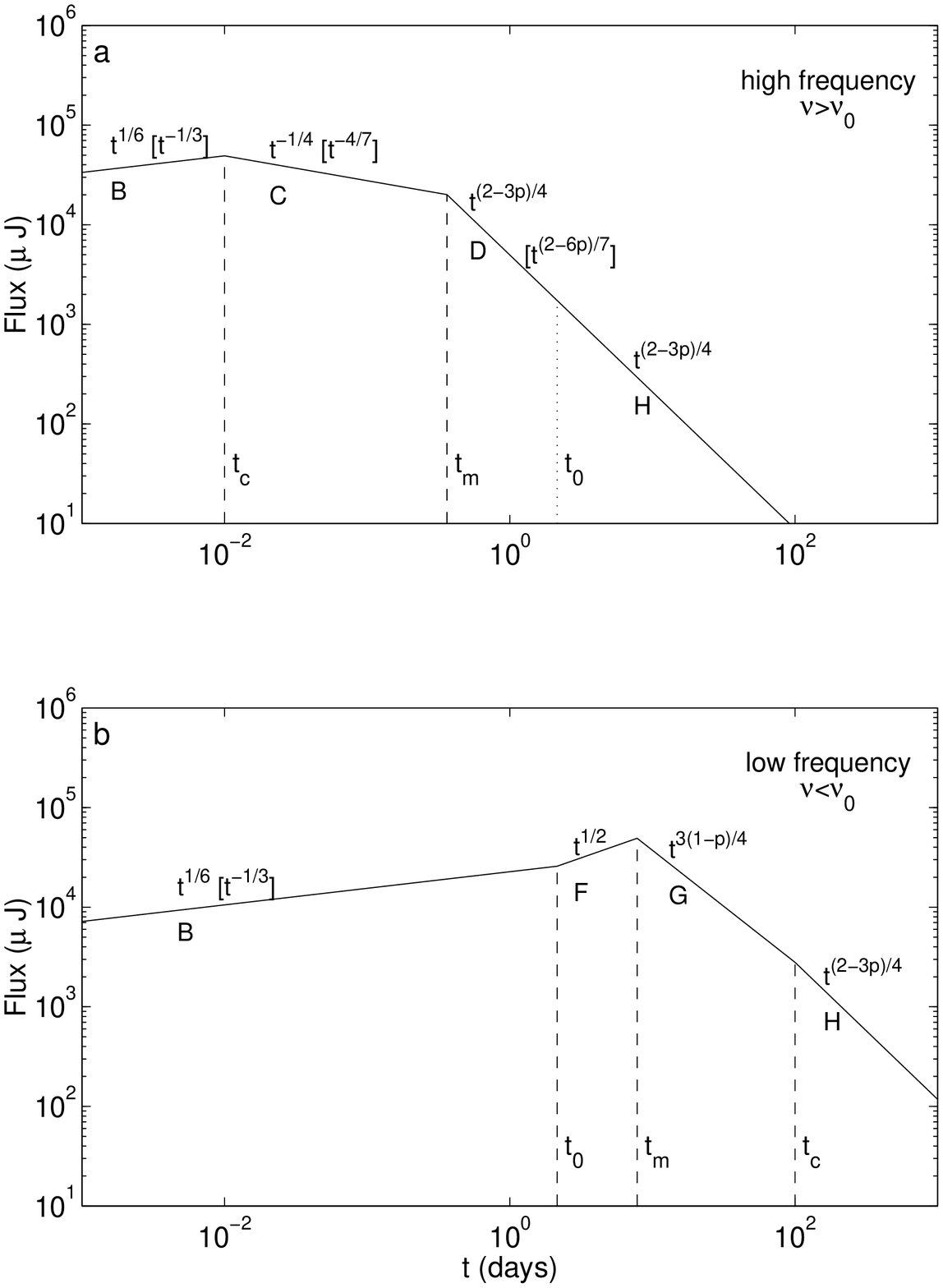,height=5.0in}}
\caption[]{
Theoretical lightcurves corresponding to the afterglow models shown in Fig. 1,
in the high frequency (a) and low frequency (b) regimes.  The four segments
that are separated by the critical times as labeled correspond to the spectral
segments in Fig. 1.  The observed flux varies with time as indicated; the
scalings within square brackets are for radiative evolution, and the other
scalings are for adiabatic evolution. 
From \cite{SPN98}.
}
\end{figure}

The basic model assumes that electrons are accelerated by the shock into a
power-law distribution $N(\gamma _{e})\sim \gamma _{e}^{-p}$ for $\gamma
_{e}>\gamma _{m}$. The lower cutoff of this distribution is assumed to be a
fixed fraction of equipartition. It is also assumed that a considerable
magnetic field is being built behind the shock, it is again characterized by a
certain fraction $\epsilon _{B}$ of the equipartition.  The relativistic
electrons then emit synchrotron radiation which is the observed afterglow.  
The broad band spectrum of such emission was given by Sari, Piran \& Narayan
\cite{SPN98} (see Figure 1). 

At each instant, there are three characteristic frequencies: 
\begin{enumerate}
\item[($i$)] $\nu_{m}$ which is the synchrotron frequency of the minimal energy
             electrons, having a Lorentz factor $\gamma _{m}$. 
\item[($ii$)] The cooling time of an electron is inversely proportional to its
             Lorentz factor $\gamma _{e}$. Therefore, electrons with a Lorentz 
             factor higher than a critical Lorentz factor 
             $\gamma _{e}>\gamma _{c}$ 
             can cool on the dynamical timescale of the system.  This 
             characteristic Lorentz factor corresponds to the 
             ``cooling frequency'' $\nu_{c}$. 
\item[($iii$)] Below some critical frequency $\nu _{a}$ the flux is self
             absorbed and is given by the Rayleigh-Jeans portion of a 
             black body spectrum. 
\end{enumerate}

The evolution of this spectrum as a function of time depends on the
hydrodynamics.  The simplest, which also describes the data well, is the
adiabatic model with a constant density surrounding medium.  The rest mass
collected by the shock at radius $R$ is about $R^{3}\rho$.  On the average, the
particles move with a Lorentz factor $\Gamma ^{2}$ in the observer frame,
and therefore the total energy is given by $E\sim \Gamma ^{2}R^{3}\rho c^{2}$.
Assuming that the radiated energy is negligible compared to the flow energy, we
obtain that $\Gamma \sim R^{-3/2}$ or in terms of the observer time,
$t=R/\Gamma^2c$, we get $\Gamma \sim t^{-3/8}$.  If the density drops as
$R^{-2}$ (as is expected if the surrounding is a wind produced earlier by the
progenitor of the burst) we get $\Gamma \sim t^{-1/4}$.  These simple scaling
laws lead to the spectrum evolution (see Figure 2). 

One can then construct light curves at any given frequency.  These will consist
of power laws, changing from one power law to the other once the break
frequencies sweep through the observed band. These power laws are in very good
agreement with the observations. 

\begin{figure}
\centerline{\psfig{figure=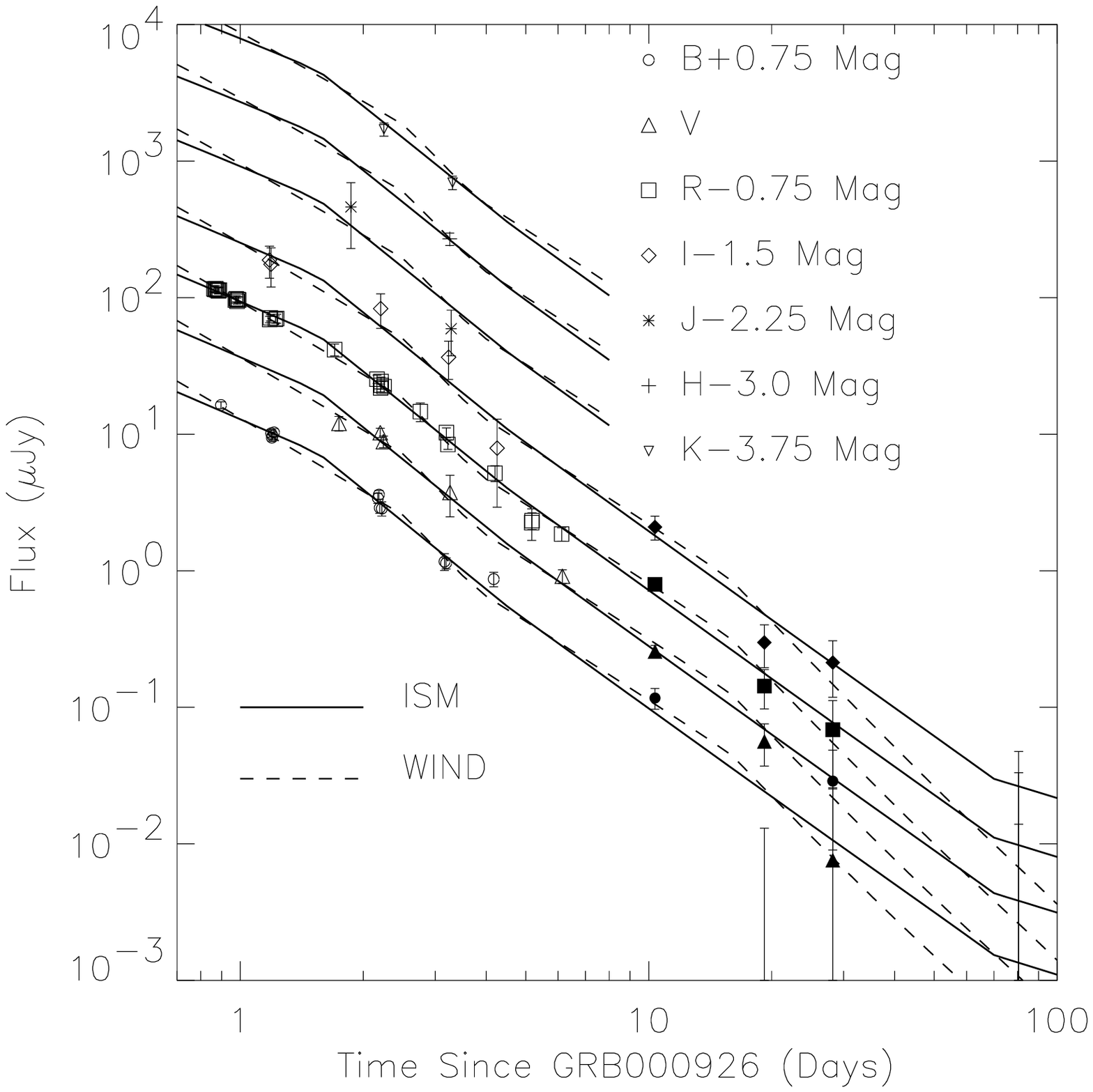,height=3.0in}}
\caption[]{
The $BVRIJHK$ light-curves of the afterglow of GRB~000926, from 8 hours to 80
days after the GRB, with model fits corresponding to an isotropic ISM and an
$\rho \sim r^{-2}$ (i.e., simple stellar wind) medium \cite{har+01}.  Also
evident is a break in the light-curve at $t \sim 1.5$ days, interpreted as
evidence for collimation of the ejecta \cite{price01}.  Fluxes from the
underlying host galaxy and another contamination galaxy have been subtracted
using late-time HST observations. 
}
\end{figure}

Observations of GRB afterglows span a broad range of frequencies, roughly from
$\sim 1$ GHz to $\sim 10^9$ GHz, and time scales from hours (or even minutes)
to a few years after the burst.  Their broad-band, time-dependent modelling,
as illustrated, e.g., in Figs. 3 and 4,
can provide considerable insights into the physics and geometry of afterglows
and even the nature of the progenitors. 

Afterglows typically have energies $< 10^{51}$ erg, and power-law electron
energy distributions with index $p \approx 2.3$.  However, some afterglows
appear to have harder electron energy distributions, with $p \approx 1.5$ 
and a high energy cutoff \cite{pk00b,gea01,pap+01}.  It is currently difficult
to distinguish between models in which the ejecta expand into a 
$\rho \sim r^{-2}$ wind-stratified ISM, and models in which the ejecta expand
into a $\rho \sim const.$ ISM, although the latter appears to be preferred in
some cases \cite{har+01}.  Early-time measurements may help distinguish between
these possibilities.  There have been suggestions \cite{pk00b} that the 
particle density of the ISM can be very low in some cases, 
$\sim 10^{-3} - 10^{-4}$ cm$^{-3}$.
While this may present difficulties for the collapsar-type models, it might be
explained in terms of pre-existing superbubbles \cite{sw01}. 

\begin{figure}
\centerline{\psfig{figure=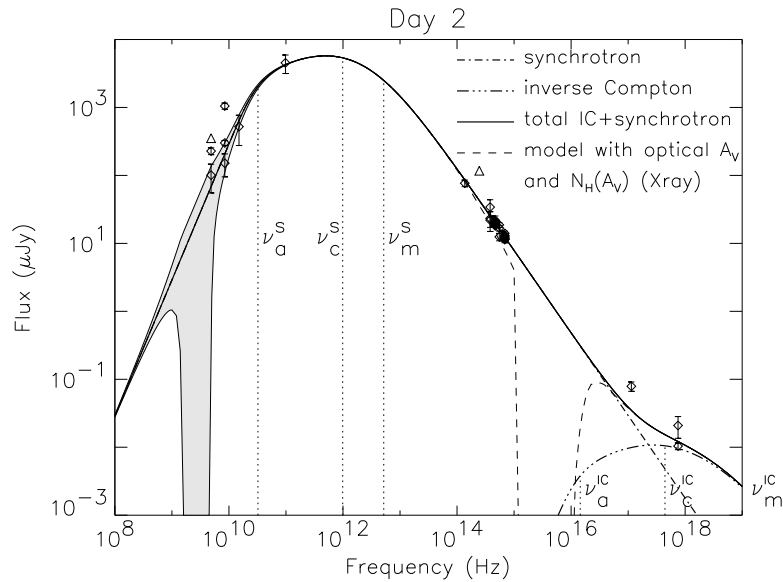,height=3.0in}}
\caption[]{
Broad-band spectrum fit to observations of the bright afterglow of GRB~000926
two days after the burst \cite{har+01}.  Proper fitting required treatment of
the interstellar scintillation (grey envelope at low frequencies), host
extinction in the optical (dotted line) and inverse-Compton scattering which
caused the excess in the X-rays. 
}
\end{figure}

Broad-band modelling is complicated by several effects:
\begin{itemize}
\item Interstellar scintillation in the radio lightcurve
\item Extinction in the optical-NIR (rest-frame UV) which has the
      primary effect of altering the observed optical spectral slope
\item The possibility of inverse-Compton scattering, which can dominate the
      electron cooling and can also produce excess emission in the X-ray 
\item The presence of a host galaxy, which may obscure the afterglow
      evolution
\end{itemize}
However, each of these complications can also provide useful additional
information.  Interstellar scintillation can be used to super-resolve
the afterglow \cite{fkn+97}.
Extinction can be used to probe the environment of the 
GRB \cite{price01,lee+01}. 
Detection of inverse-Compton emission gives a better handle on the
density of the medium \cite{har+01}.
Detection of the host galaxy in the radio \cite{bkf01} or sub-mm \cite{fr+01b}
can give more complete estimates of the star-formation rate, while optical host
studies can also give clues to the nature of the progenitors \cite{bkd01}.

Radio observations in particular provided the first direct evidence for
relativistic motions in GRBs, through the use of interstellar scintillation
to measure the physical expansion rate of an afterglow \cite{fkn+97,fwk00}.

Another interesting phenomenon was the detection of a bright (9th magnitude),
prompt optical emission simultaneous with GRB 990123 \cite{A99}. 
Theoretical explanation for such a flash \cite{SP99a,SP99b,MR97} is that
there are initially two shocks in a GRB: a forward shock going into the 
surrounding medium, and a reverse shock going into the expanding shell. 
This model could neatly account for the observed optical properties of 
GRB 990123.  It takes tens of seconds for the reverse shock to sweep through
the ejecta and produce the bright flash.  Later, the shocked hot matter expands
adiabatically and the emission quickly shifts to lower frequencies and
considerably weakens.  Another new ingredient that was found in GRB 990123 is a
radio flare \cite{kfs+99}.  In most afterglows the radio peaks around few weeks
and then decays slowly, but this burst had a fast rising flare, peaking around
a day and decaying quickly.  The optical flash and the radio flare are closely
related.  Similar radio flares were detected from a few other bursts as well.

\section{Host Galaxies and Redshifts}

Host galaxies of GRBs serve a dual purpose:  First, in most cases redshifts 
are measured for the host galaxy, rather than the afterglow itself (sometimes
both).  This is mainly because most optical afterglows so far have been
discovered too late for an effective absorption-line spectroscopy, but also
in some cases no optical transient (OT) is detected, but a combination of
the X-ray (XT) and radio transient (RT) unambiguously pinpoints the host
galaxy.  Second, properties of the hosts and the location of OTs within them
can provide valuable clues about the possible nature of the progenitors,
e.g., their relation to the massive star formation, etc.

\begin{figure}
\centerline{\psfig{figure=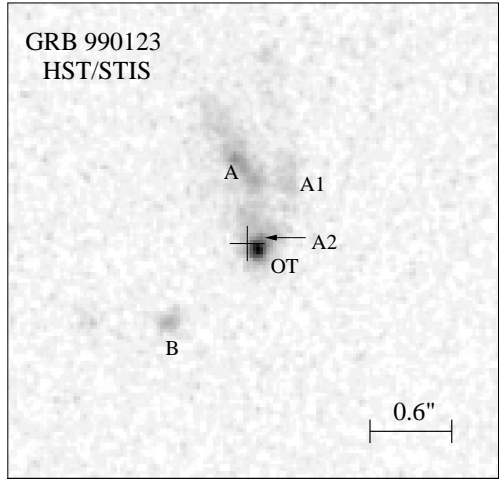,height=2.5in}}
\caption[]{
Image of the host galaxy of GRB 990123 at $z = 1.600$, obtained with the 
HST \cite{bod+99}.
The cross marks the position of the optical afterglow, from the ground-based 
measurement.  The host is an irregular, possibly merging system.
}
\end{figure}

\begin{figure}
\centerline{\psfig{figure=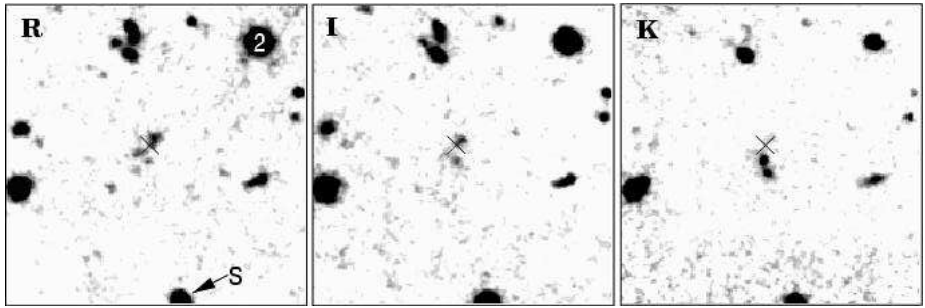,width=4.8in,angle=270}}
\caption[]{
The host galaxy system of GRB 980613 in $R$ (left), $I$ (middle) and $K$ bands
(right), from images obtained at the Keck \cite{sgd+01a}.
The ``$\times$'' marks the location of the OT.  Note the complex morphology
of the system, suggestive of mergers, and a variety of colors among the
galaxy components.
}
\end{figure}

Table 1 summarizes the host galaxy magnitudes and redshifts known to us as
of mid-June 2001.  The median apparent magnitude is $R = 24.8$ mag, with
tentative detections or upper limits reaching down to $R \approx 29$ mag.
Down to $R \sim 25$ mag, the observed distribution is consistent with deep
field galaxy counts \cite{shyc95}, but fainter than that, selection effects may
be playing a role.  We note also that the observations in the visible probe the
UV in the restframe, and are thus especially susceptible to extinction. 

\begin{center}
\begin{table}[t]

\caption{\centerline {\bf GRB Host Galaxies and Redshifts}}

{~ \\ ~}

\mbox{~}
\begin{tabular}{lrlll}

  GRB &    
  \multcc{$R$ mag} &    
  Redshift &    
  \multcc{Type $^1$} &   
  References \\ \hline \\

  970228     &    25.2  &   0.695    & e   & \cite{fpt+99,bdk01} \\
  970508     &    25.7  &   0.835    & a,e & \cite{mdk+97,bdkf98,fpg+00} \\
  970828     &    24.5  &   0.9579   & e   & \cite{sgd+01b} \\
  971214     &    25.6  &   3.418    & e   & \cite{kdr+98,odk+98} \\
  980326     &    29.2  & $\sim$1?   &     & \cite{bkd+99,af+01} \\
  980329     &    27.7  & $<$3.9     & (b) & \cite{sh+00c,sgd+01c} \\
  980425 $^2$&    14    &   0.0085   & a,e & \cite{gvv+98} \\
  980519     &    26.2  &            &     & \cite{hpja99} \\
  980613     &    24.0  &   1.097    & e   & \cite{sgd+01a} \\
  980703     &    22.6  &   0.966    & a,e & \cite{dkb+98} \\
  981226     &    24.8  &            &     & \cite{fkb+99} \\
  990123     &    23.9  &   1.600    & a,e & \cite{kdo+99,bod+99} \\
  990308 $^3$& $>$28.5  &            &     & \cite{sh+00} \\ 
  990506     &    24.8  &   1.30     & e   & \cite{tbf+00,bfs01} \\
  990510     &    28.5  &   1.619    & a   & \cite{af+00b,vfk+01} \\
  990705     &    22.8  &   0.86     & x   & \cite{mpp+00,pgg+00} \\
  990712     &    21.8  &   0.4331   & a,e & \cite{hhc+00a,hhc+00b,vfk+01} \\
  991208     &    24.4  &   0.7055   & e   & \cite{csg+01,sgd+99c} \\
  991216     &    24.85 &   1.02     & a,x & \cite{pmv+00,pmv+99,pgg+00} \\
  000131     & $>$25.7  &   4.50     & b   & \cite{ahp+00} \\
  000214     &          & 0.37--0.47 & x   & \cite{apv+00} \\
  000301C    &    28.0  &   2.0335   & a   & \cite{jfg+01,as+00,smc+00,fv+01} \\
  000418     &    23.9  &   1.1185   & e   & \cite{mm+00,jsb+00} \\
  000630     &    26.7  &            &     & \cite{dlk+01} \\
  000911     &    25.0  &   1.0585   & e   & \cite{pap+01} \\
  000926     &    23.9  &   2.0369   & a   & \cite{jpuf+00,jpuf+00b,smc+00b} \\
  010222     & $>$24    &   1.477    & a   & \cite{sh+01,pmg+01,jsb+01,jpg+01} \\

\\ 
\hline
\end{tabular}

\mbox{~}\\

{\small
\textsc{Notes}: \\
$^1$ e = line emission, a = absorption, b = continuum break, x = x-ray \\
$^2$ Association of this galaxy/SN/GRB is somewhat controversial \\
$^3$ Association of the OT with this GRB may be uncertain \\
}
\end{table}
\end{center}

\begin{figure}
\centerline{\psfig{figure=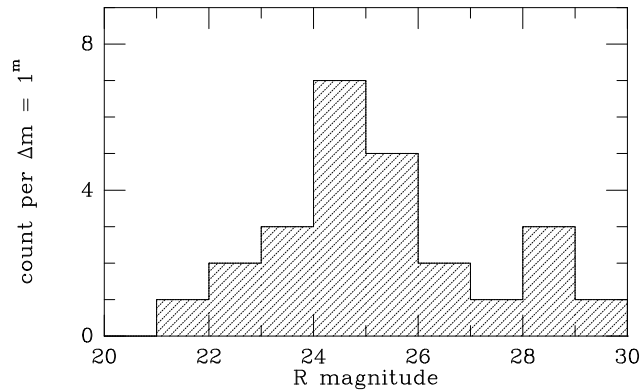,height=2.0in}}
\caption[]{
Histogram of the $R$ band magnitudes of GRB host galaxies as of June 2001.  
The decline at $R > 25$ mag may be due at least in part to the observational
selection effects.  The median for this sample is at $R = 24.8$ mag.
}
\end{figure}

\begin{figure}
\centerline{\psfig{figure=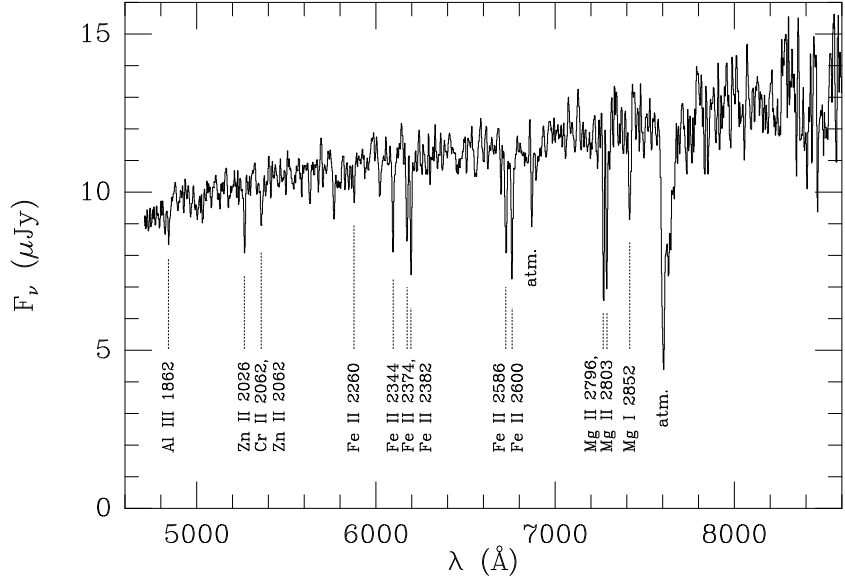,height=3.0in}}
\caption[]{
Spectrum of the OT associated with GRB 990123, obtained at the 
Keck \cite{kdo+99}.  The prominent absorption lines correspond to the host
galaxy redshift, $z = 1.600$.
}
\end{figure}

\begin{figure}
\centerline{\psfig{figure=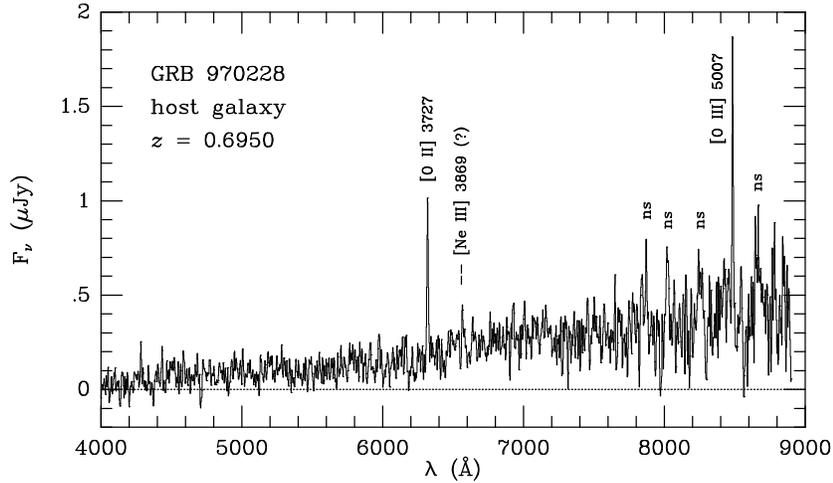,height=2.5in}}
\caption[]{
Spectrum of the host galaxy of GRB 970228, obtained at the Keck \cite{bdk01}. 
Prominent emission lines [O II] 3727 and [O III] 5007 and possibly [Ne III]
3869 are labeled assuming the lines originate from the host at redshift $z =
0.695$.  The notation ``ns'' refers to Noise spikes from strong night sky
lines are labeled ``ns''.
}
\end{figure}

\begin{figure}
\centerline{\psfig{figure=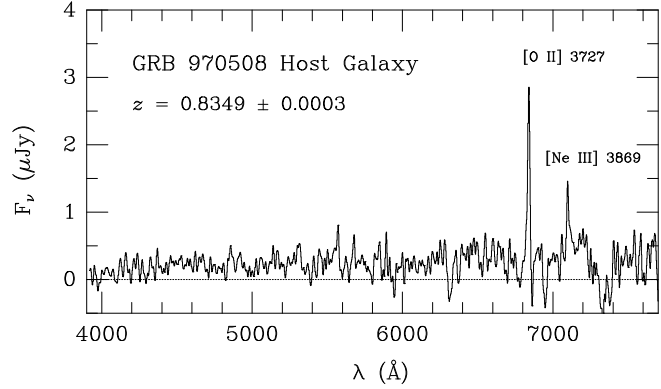,height=2.0in}}
\caption[]{
Spectrum of the host galaxy of GRB 970508, obtained at the Keck \cite{bdkf98}.
Note the strong [Ne III] line, indicative of the high-temperature H II regions,
presumably powered by UV radiation from massive stars.
}
\end{figure}

\begin{figure}
\centerline{\psfig{figure=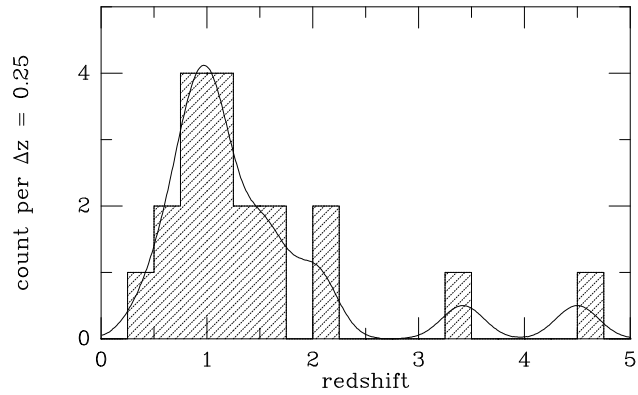,height=2.0in}}
\caption[]{
Histogram of the GRB redshifts as of June 2001.  The solid line is a
Gaussian-smoothed distribution.  The median for this sample is $z = 1.1$.
The decline at $z > 1.3$ or so can due in part to the observational selection
effects (both due to the more distant galaxies being fainter, and the absence
of strong emission lines for spectra at $z \sim 1.3 - 2$). 
}
\end{figure}

\begin{figure}
\centerline{\psfig{figure=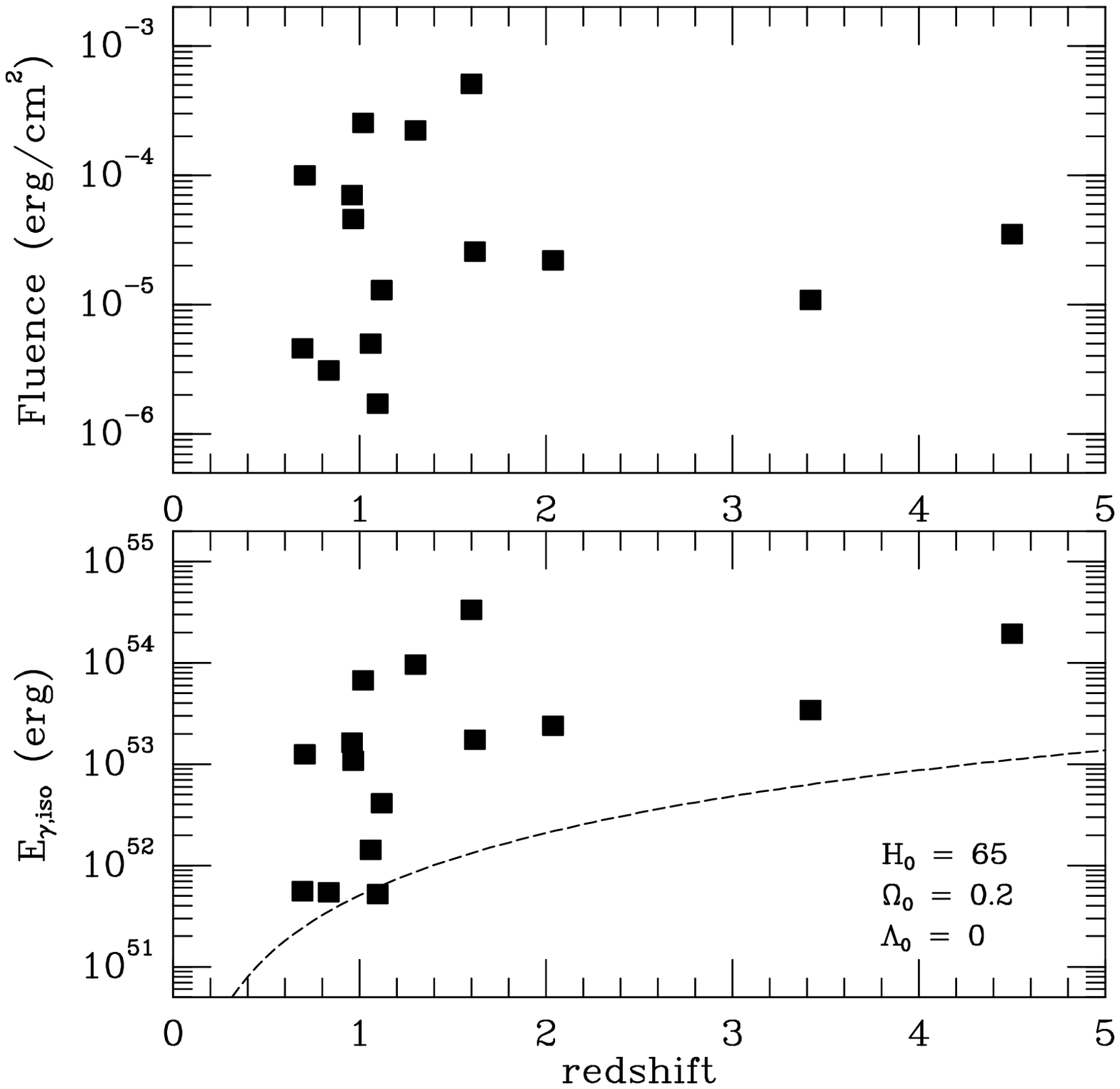,height=3.0in}}
\caption[]{
Observed $\gamma$-ray fluence (top) and the corresponding isotropic
$\gamma$-ray energy (bottom), for the bursts with measured redshifts and
published fluences, as of June 2001.  Note the complete lack of correlation
between the fluence (roughly proportional to the mean flux) and the redshift.
In computing the energies, a simple Friedmann cosmology with $H_0 = 65$ km
s$^{-1}$ Mpc$^{-1}$, $\Omega_0 = 0.2$, and $\Lambda_0 = 0$ was used.  The
dashed line in the bottom panel corresponds to the limiting fluence of
$2 \times 10^{-6}$ erg cm$^{-2}$.  It is clear that even with the present
generation of instruments, we can in principle detect bursts out to very
high redshifts.  For a more detailed discussion of k-corrected, bolometric
fluxes and energies, see \cite{bfs01}.
}
\end{figure}

The redshift distribution of GRB hosts to date (Fig. 11)
is about what is expected for an evolving, normal field galaxy population at
these magnitude levels.  There is an excellent qualitative correspondence
between the observations and simple galaxy evolution models \cite{maomo98}. 
The majority of redshifts so far are from the spectroscopy of host galaxies,
and some are based on the absorption-line systems seen in the spectra of the
afterglows (which are otherwise featureless power-law continua).  Reassuring
overlap exists in some cases; invariably, the highest-$z$ absorption system
corresponds to that of the host galaxy, and has the strongest lines.
A new method for obtaining redshifts may come from the X-ray spectroscopy of
afterglows, using the Fe K line at $\sim 6.55$ keV \cite{pcf+99,apv+00,pgg+00},
or the Fe absorption edge at $\sim 9.28$ keV \cite{yno+99,wmkr00,amati00}.
In general, rapid X-ray spectroscopy of GRB afterglows may become a powerful
tool to understand their physics and origins. 

Almost all GRB redshifts measured to date required large telescopes (this is
certainly true for the measurement of host galaxy redshifts), although it is
certainly possible to measure absorption redshifts of OTs, if they are
identified quickly \cite{jpg+01}.
The redshifts are still in a short supply, and there has been a considerable
interest in trying to produce a photometric redshift estimator for GRBs from
their $\gamma$-ray light curves alone \cite{sps99,fr00,nmb00,rlf+01} (Fig. 13).
While rough redshifts (perhaps good to a factor of 2?) can be predicted,
the practical utility of these relations remains to be tested.

\begin{figure}
\centerline{\psfig{figure=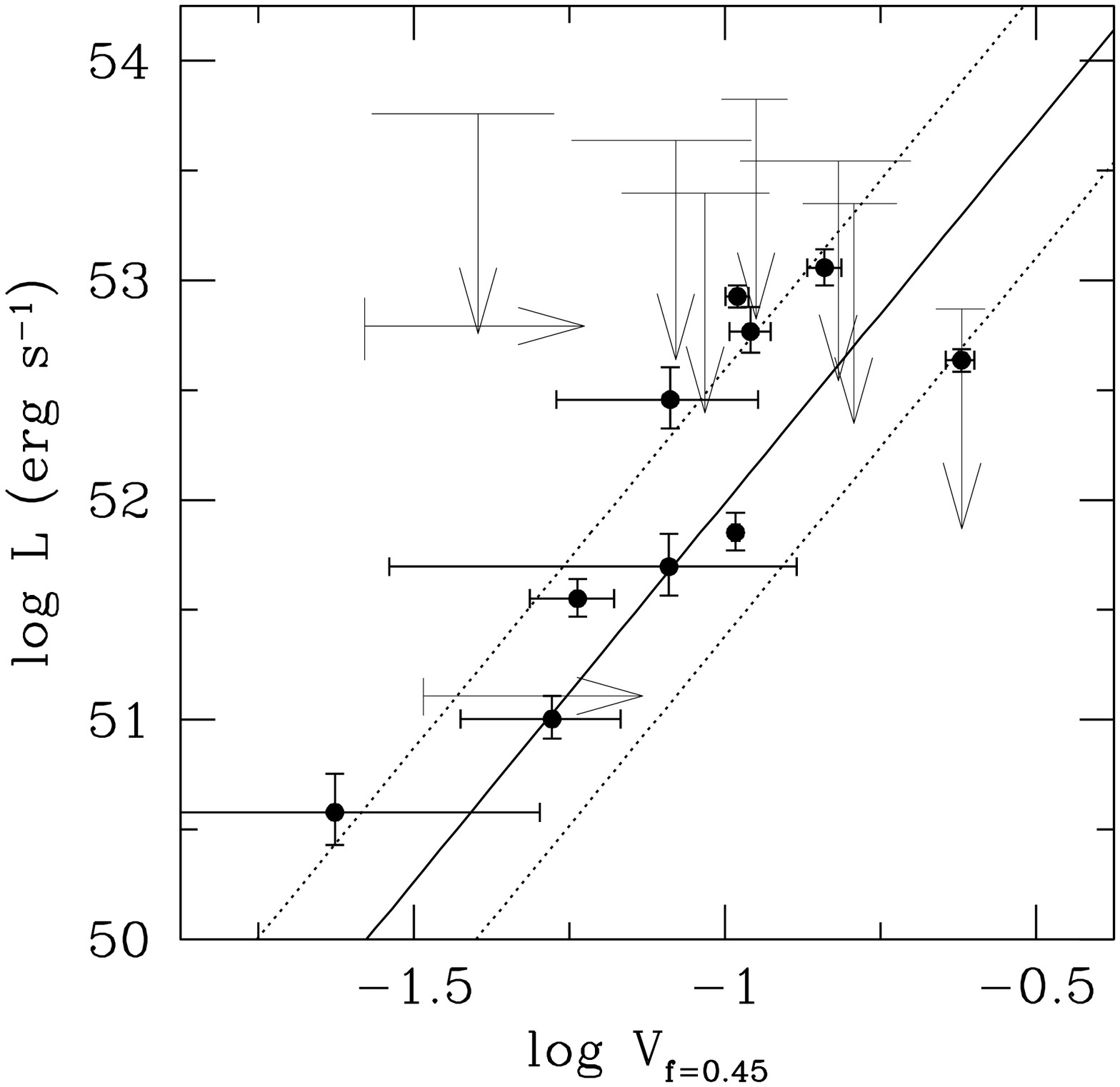,height=2.5in}}
\caption[]{
A correlation between the apparent isotropic $\gamma$-ray energy vs. a
variability parameter, from \cite{rlf+01}.  Solid and dashed lines represent
the best fit power-law $\pm 1~\sigma$.  The fit includes bursts with actual
measurements, and upper limits.
}
\end{figure}

Returning to the issue of GRB host galaxies, the first question is, are they
special in some way?  Their magnitude and redshift distributions are typical
for the normal, faint field galaxies, as are their morphologies \cite{hol01}
when observed with the HST: often compact, sometimes suggestive of a merging 
system \cite{sgd+01a}, but that is not unusual for galaxies at comparable
redshifts.

If GRBs are somehow related to the massive star formation 
(e.g., \cite{tot97,pac98b}, etc.), 
it may be worthwhile to examine their absolute luminosities and star formation
rates (SFR), or spectroscopic properties in general.  This is hard to 
answer \cite{kth98,hf99,sch00}
from their visible ($\sim$ restframe UV) luminosities alone: the observed 
light traces an indeterminate mix of recently formed stars and an older
population, cannot be unambiguously interpreted in terms of either the total
baryonic mass, or the instantaneous SFR.  

Fig. 14
shows the distribution of absolute $B$-band luminosities of GRB hosts
identified to date, under two extreme assumptions about evolutionary
corrections.  The hosts appear to be somewhat subluminous relative to a
present-day average ($L_*$) galaxy.  This is generally expected, since much of
the SFR activity at $z \sim 0.5 - 1$ (and probably beyond) among the
field galaxies appears to be in subluminous systems \cite{lth+95,ellis97}, 
and thus the GRB hosts would be representative of the normal, star-forming
field galaxy population.  One could also speculate that the trend towards
lower luminosity galaxies may be really selecting on average lower metallicity
systems, where the mean extinction may be lower, making them easier to detect,
or whose stellar IMF may be biased towards more massive stars (this is highly
speculative).  A much larger sample of GRB hosts is needed in order to better
understand this issue. 

\begin{figure}
\centerline{\psfig{figure=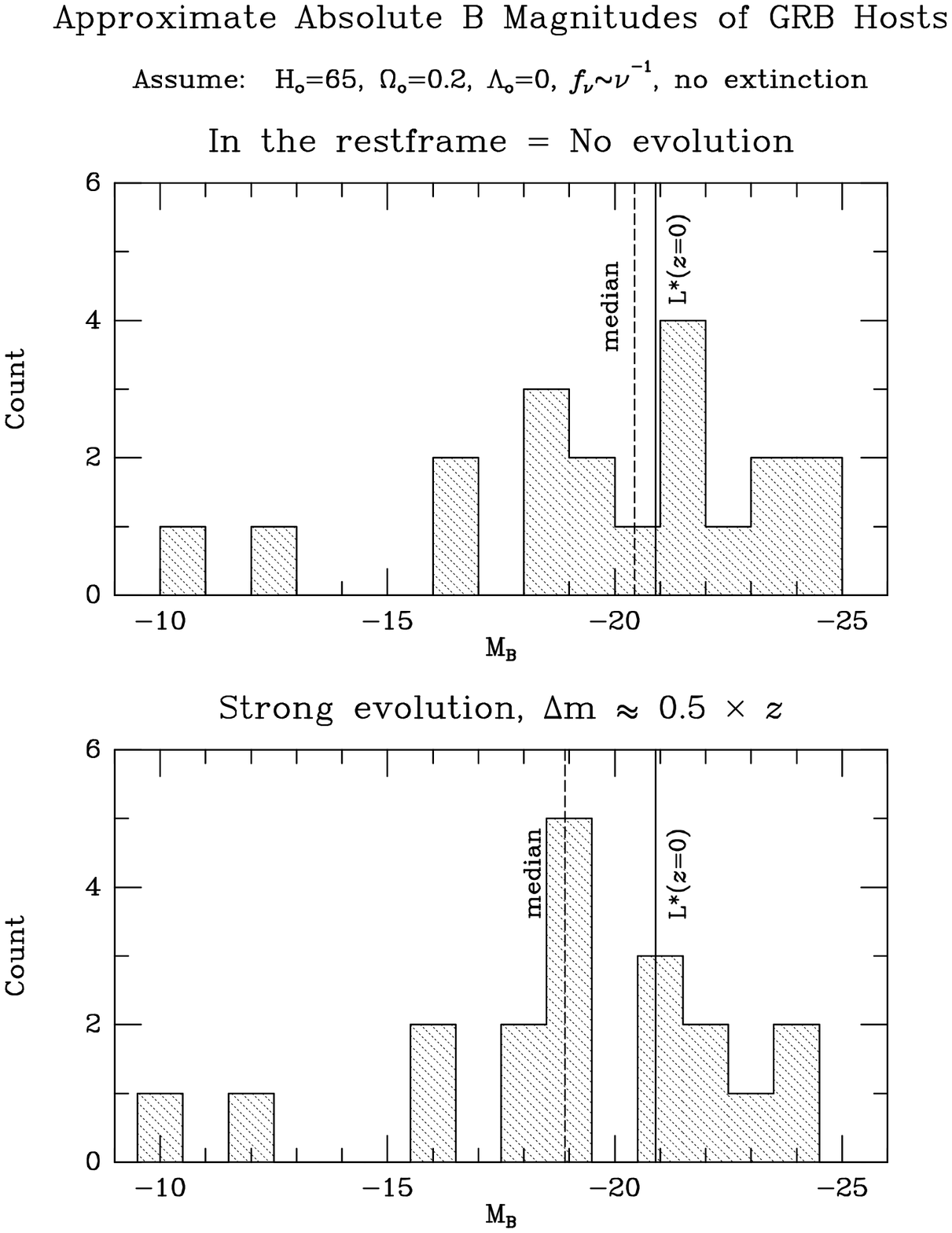,height=4.5in}}
\caption[]{
Distribution of estimated absolute $B$-band magnitudes for GRB host galaxies
with known redshifts, as of June 2001.  These restframe magnitudes were
computed from the observed $R$-band magnitudes by approximating the galaxy
spectra as $f_\nu \sim \nu^{-1}$ and no additional extinction correction. 
Standard Friedmann cosmology with $H_0 = 65$ km s$^{-1}$ Mpc$^{-1}$, $\Omega_0
= 0.2$, and $\Lambda_0 = 0$ was used (the results are not strongly sensitive to
these assumptions).  The top panel shows the magnitudes as they are observed,
i.e., assuming no evolution from the observed host redshift to the present day.
The bottom panel incorporates a simple evolutionary fading correction (i.e.,
galaxies were assumed to have been brighter in the past, and so they would be
fainter today), as $\Delta m = 0.5 z$; this is about as strong as suggested by
the modern field galaxy evolution studies.  Solid lines indicate the values of
$M_B$ corresponding to an $L_*$ galaxy today; dashed lines are the sample
medians. 
}
\end{figure}

Spectroscopic measurements provide direct estimates of the recent, massive
SFR in GRB hosts.  Most of them are based on 
the luminosity of the [O II] 3727 doublet \cite{ken98}, 
the luminosity of the UV continuum at $\lambda_{rest} = 2800$ \AA\ \cite{mpd98},
in one case so far \cite{kdr+98}
from the luminosity of Ly$\alpha$ 1216 line \cite{tdt95},
and in one case \cite{dkb+98}
from the luminosity of Balmer lines \cite{ken98}.
All of these estimators are susceptible to the internal extinction and its
geometry, and have an intrinsic scatter of at least 30\%.
The observed SFR's range from a few tenths to a few $M_\odot$ yr$^{-1}$,
again typical for the normal field galaxy population at comparable redshifts.

Equivalent widths of the [O II] 3727 doublet in GRB hosts, which may provide
a crude measure of the SFR per unit luminosity (and a worse measure of the
SFR per unit mass), are on average somewhat higher \cite{sgd+01c}
than those observed in magnitude-limited field galaxy samples at comparable
redshifts \cite{hcbp98}.
Another intriguing hint comes from the flux ratios of [Ne III] 3869 to
[O II] 3727 lines: they are on average a factor of 4 to 5 higher in GRB
hosts than in star forming galaxies at low redshifts.  These strong [Ne III] 
require photoionization by massive stars in hot H II regions, and may represent
an indirect evidence linking GRBs with massive star formation.

The interpretation of the luminosities and observed star formation rates is
vastly complicated by the unknown amount and geometry of extinction.  The
observed quantities (in the visible) trace only the unobscured stellar
component, or the components seen through optically thin dust.  Any 
stellar and star formation components hidden by optically thick dust cannot
be estimated at all from these data, and require radio and sub-mm observations
(see below).
Thus, for example, optical colors of GRB hosts cannot be used to make any
meaningful statements about their net star formation activity.  The broad-band
optical colors of GRB hosts are not distinguishable from those of normal
field galaxies at comparable magnitudes and redshifts \cite{bdk01,sfct+01}.

A great deal about the progenitors can be gleaned from the locations of GRBs in
and around galaxies.  GRBs from collapsars are expected to occur where massive
stars are formed, in molecular clouds and HII regions, whereas GRBs from
degenerate binaries that merge after at least one SN in the system (e.g.,
NS--NS and NS--BH binary systems) can merge a few to hundreds of kiloparsecs
from their birthsite.  This is because of the large systemic kicks imparted to
binary systems following a SN explosion.  Modeling of the location of merging
binaries suggests that $\sim 1/3 - 1/2$ of all GRBs should occur $> 10$ kpc in
projection from the centers of their host galaxies \cite{bsp99,fwh99,bbz00}.

\begin{figure}
\centerline{\psfig{figure=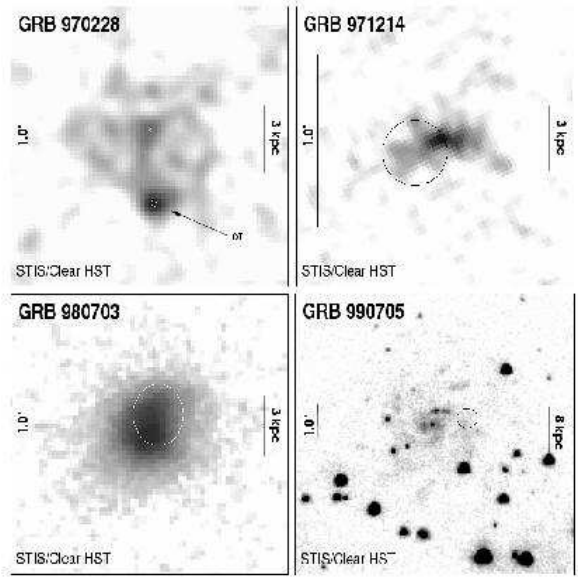,height=3.0in}}
\caption[]{
Example GRB-host galaxy offsets, from \cite{bkd01}.
The 3-$\sigma$ error contours of the GRB afterglow locations are shown as
ellipses.  In the top left panel, the afterglow of GRB 970228 is still visible
the the south of the center of the host.  In the other three, the afterglow has
faded from view by the epoch of the respective HST observations. 
}
\end{figure}

The proximity of OTs associated with GRB 970228 and GRB 970508 to their hosts
already posed a difficulty for the merger scenario \cite{bdkf98,pac98b}.
We completed the first comprehensive study of the offset location of 20 
well-localized cosmological GRBs using extensive ground-based and space-based 
imgaging data \cite{bkd01}.  Fig. 15
shows some examples of GRB-host offsets.  The 3-$\sigma$ error contour of the
GRB location is marked as well as a physical scale for each galaxy.  We find
that all well-localized GRBs to-date fall within 1.2 arcsec ($\sim 10$ kpc in
projection) from the nearest detected galaxy. Statistically, most if not all of
these nearby host galaxies are indeed physically associated with the respective
GRB and not spurious superpositions: the chance of superposition of more than
3 GRB hosts is $< 1$\%. 

\begin{figure}
\centerline{\psfig{figure=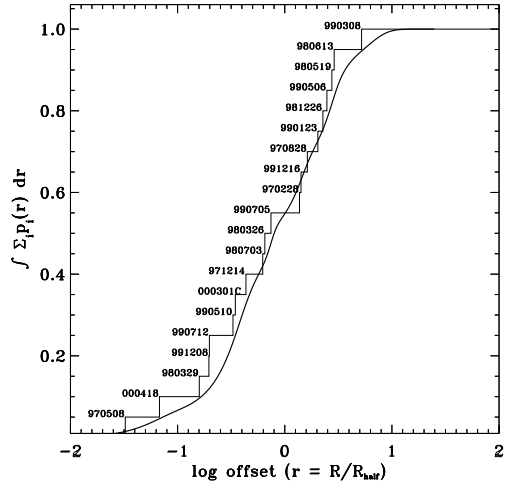,height=2.5in}}
\caption[]{
Cumulative offset distribution of 20 GRBs about their respective host galaxies.
The distance measure ($r$) is in units of the host half-light radius.  
The distribution is consistent with a simple exponential disk model (an
approximation to the location of massive star formation in late-type galaxies)
with a Kolmogorov-Smirnov (KS) probability of 0.18.  The smoothed solid line is
a representation of the cumulative histogram which accounts for the
uncertainties in the individual measurements. 
}
\end{figure}

The distribution of GRB-host offsets closely follows the light of their hosts,
which is roughly proportional to the density of star formation (especially for
the high-$z$ galaxies).  It is thus fully consistent with a progenitor
population associated with the sites of massive star formation.
Fig. 16
shows the observed offsets distribution compared with a simple exponential disk
model for the distribution of massive star formation.  The observed 
distribution also appears to be inconsistent with the merger scenario, where
progenitors travel far from their birthsites.

\section{The GRB-Supernova Connection}

A direct consequence of the collapsar model is that GRBs are expected
to be accompanied by supernovae (SNe).
The first evidence for a possible GRB/SN connection was provided by
the discovery of SN\,1998bw in the error box of GRB 980425
\cite{gvv+98}.  The temporal and spatial coincidence of SN\,1998bw with
GRB 980425 suggest that the two phenomena are related
\cite{gvv+98,kfw+98}; however, the actual identification of the SN with
the X-ray afterglow remains somewhat uncertain \cite{paa+00}.  
An additional indication that SN\,1998bw may be related to GRB 980425 comes
from the fact that the radio emitting shell in SN\,1998bw must be expanding at
relativistic velocities, $\Gamma \geq 2$ \cite{kfw+98}, which was previously
never observed in a SN.  From minimum energy arguments, it was estimated that
this relativistic shock carried $5\times 10^{49}$ erg, and could well have
produced the GRB at early time.  Further, detailed analysis of the radio light
curve \cite{cl99} showed additional energy injection one month after the SN
event -- highly suggestive of a central engine (i.e., a BH vs. NS formation)
rather than a purely impulsive explosion. 

\begin{figure}
\centerline{\psfig{figure=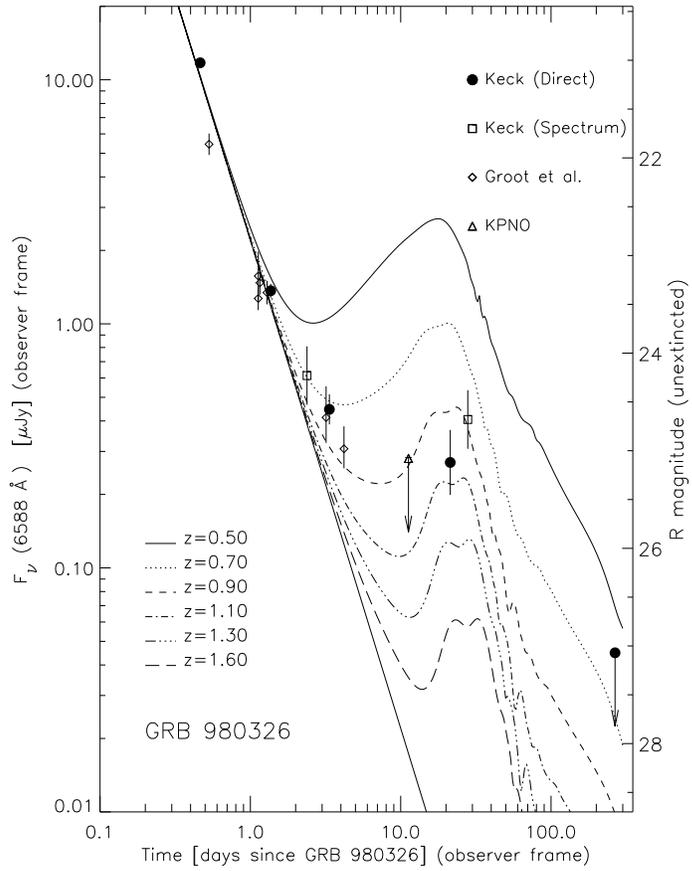,height=4.5in}}
\caption[]{
R-band light curve of GRB 980326 and the sum of an initial power-law
decay plus SN Ic light curve for redshifts ranging from $z = 0.50$ to $1.60$. 
The best-fit redshift for the SN is $z \approx 0.95$, and it is consistent 
with the observed spectrum taken during the reburst.  From \cite{bkd+99}.
}
\end{figure}

If this identification is correct, GRB 980425 is most certainly not a typical
GRB: the redshift of SN\,1998bw is 0.0085 and the corresponding $\gamma$-ray
peak luminosity of GRB 980425 and its total $\gamma$-ray energy budget are
about a factor of $\sim$ 10$^5$ smaller than those of ``normal'' GRBs
\cite{gvv+98}.  But if the identification is correct, such SN-GRBs may well be
the most frequently occurring GRBs in the universe. 

A probable association of a ``normal'' GRB with a SN-like event was the
discovery \cite{bkd+99}
of the late-time rebrightening (with a much redder spectrum than that of the
early-time afterglow) of GRB 980326 (see Fig. 17).
Similar deviations in the light curve and colors were also seen in the case of
GRB 970228, with the same physical interpretation \cite{ggv+97,rei99,gtv+00}.
However, it should be noted that many other GRB aftergrlows did $not$ show
such lightcurve deviations, even if they were detectable in principle.
An alternative explanation for this phenomenon involves light 
echoes \cite{eb00,rei01b}.

\section{Jets and Beaming}

The question of jets and beaming in GRBs has been discussed for some years, but
it was really brought into focus when the redshift measurements
\cite{kdr+98,kdo+99} 
of GRB 971214 and GRB 990123, implied the isotropic $\gamma$-ray energy
releases approaching $\sim 10^{54}$ erg.  If GRBs are collimated, there are two
important implications: 
First, the true total energy emitted by the source is smaller by a factor of
$\Omega /4\pi \sim \theta_0^2/4 $ than if the ejecta was spherical. 
Second, the true event rate must be bigger by the same factor to account for
the observed rate. 

The first attempts to constrain the beaming in GRBs were through searches
for the so-called ``orphan afterglows'', corresponding to GRBs beamed away 
from us.  The basic idea is that the relativistic beaming produces a
visibility cone with an opening angle $\sim 1/\Gamma$.  In the early
phases of a GRB, $\Gamma \sim 100$ and the $\gamma$-ray emission would
be highly beamed.  As the afterglow spectrum evolves towards the lower
frequencies, and as $\Gamma$ decreases, afterglow emission at longer
wavelengths (first X-ray, then optical, then radio) would be progressively
less beamed \cite{rho97}.  One could thus see numerous weakly beamed afterglows
without accompanying GRBs.

Several searches have been made in X-rays \cite{vik98,grin99,gre+00}, 
but no convincing population of orphan afterglows has been found, and giving
(weak) limits on the beaming factor $f_b > 10^{-3}$ or so.  Comparable limits
have been found for the existing radio surveys \cite{pl98}.  One
serendipitously found candidate has been reported in the optical \cite{djo01}. 

A more convincing evidence that GRB fireballs are not spheres but rather have
conical (jet-like) geometries comes from their light curves.  The observational
signature of conical geometry manifests itself as a panchromatic ``break'' in
the power law decay of the afterglow emission, which declines more rapidly
relative to a spherical case \cite{rho99,sph99}.  This is due to two effects. 
The first is an edge effect that occurs at at a time $t_j$ when the bulk
Lorentz factor of the blast wave $\Gamma$ has slowed down to
$\Gamma<\theta_j^{-1}$ (where $\theta_j$ is the opening angle of the jet).  
The second effect that also becomes important after $t_j$ is the lateral
spreading of the jet.  The ejecta, now encountering more surrounding matter,
decelerate faster than in the spherical case. 

The first claim of a jet was made for the radio afterglow of
GRB~970508, which showed deviations from the predictions of a simple
spherical adiabatic model \cite{wkf98}.  However, it was the spectacular
isotropic energy release \cite{kdo+99} of GRB 990123 -- approaching the
rest mass of a neutron star -- which emphasized the possible
importance of jets in GRBs.  A case for a jet in the afterglow of this
burst was made on the basis of a sharp break ($\Delta\alpha\geq
0.7$) in the optical afterglow and upper limits in the
radio \cite{kfs+99}.  The clearest evidence for a jet is a sharp break
over a broad range of frequencies and such a signature was seen in the
lightcurves of GRB 990510 at optical \cite{sgk+99,hbf+99} and
radio \cite{hbf+99} wavelengths and was found to be consistent with the
X-ray \cite{psa+00} light curve.  Furthermore, the detection of
polarization \cite{clg+99,wvg+99} from this event gave further credence
to the jet hypothesis: the non-spherical geometry leads to polarized
signal, from which the geometry of the jet can be
inferred \cite{gl99,sar99}.

More recently, the identification of jets has shifted from single frequency
measurements to global model fitting of joint optical, radio and X-ray 
datasets \cite{bsf+00,pk00b}.  This approach has the advantage that by
simultaneously fitting all the data, the final outcome is less sensitive to
deviations in small subsets of the data.  In addition, since the character of
the achromatic break is different above and below the peak of the synchrotron
spectrum \cite{sph99}, broad-band measurements give more robust determinations
of the jet parameters. This approach was crucial in distinguishing the jet
break for GRB 000301C \cite{bsf+00} whose decaying lightcurves exhibited
unusual variability \cite{mbb+00}, now attributed to microlensing \cite{gls00}.
Likewise, radio measurements were useful in determining $t_j$ for GRB 000418
because the light of the host galaxy masked the jet break at optical
wavelengths \cite{die01}. 

\begin{figure}
\centerline{\psfig{figure=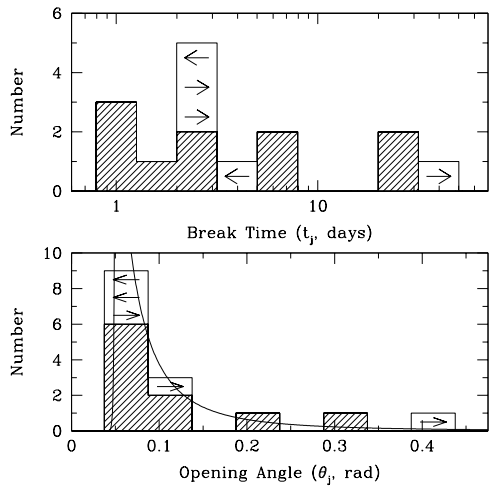,height=2.5in}}
\caption[]{
The observed distribution of light curve jet break times, $t_j$ (top), and
jet opening angles, $\theta_j$, (bottom).  A heuristic model fit (line)
assumes two power laws: $p_{\rm obs}(f_b)=(f_b/f_0)^{\alpha+1}$ for
$f_b<f_0$ and $p_{\rm obs}(f_b)=(f_b/f_0)^{\beta+1}$ for $f_b>f_0$.  Since for
every observed burst there are $f_b^{-1}$ that are not observed, the true
distribution is $p_{\rm true}(f_b) = f_b^{-1} p_{\rm obs}(f_b)$.  Fits poorly
constrain $\alpha$, and $\beta=-2.77^{+0.24}_{-0.30}$; $\log f_0
=-2.91^{+0.07}_{-0.06}$.  Thus, the true differential probability distribution
(under the small angle approximation, $f_b\propto \theta_j^2$) is given by
$p_{\rm true}(\theta_j)\propto \theta_j^{-4.54}$ with the observed distribution
being $p_{\rm obs} \propto \theta_j^{-2.54}$.  The distribution $p_{\rm
true}(f_b)$ allows us to estimate the true correction factor, $\langle
f_b^{-1}\rangle$ that has to be applied to the observed GRB rate in order to
obtain the true GRB rate.  We find $\langle f_b^{-1}\rangle =
f_0^{-1}[(\beta-1)/\beta] \sim 520\pm 85$.  From \cite{fr+01a}. 
}
\end{figure}

We have recently determined the values (or limits) of jet breaks $t_j$ for a
complete sample of all GRBs with known redshifts \cite{fr+01a}.  
Within the framework of this conical jet model \cite{sar99} we are able 
to derive the opening angle $\theta_j$.  The distribution
of the $t_j$ and $\theta_j$ is shown in Fig. 18.  
Corresponding to the wide range in $t_j$ values from $\leq 1$ d to 30 d, 
we obtain a range in $\theta_j$ from 3$^\circ$ to more than 25$^\circ$ with a
strong concentration near 4$^\circ$. This result suggests that the broad range
of fluence and luminosity observed for GRBs is largely the result of a wide
variation of opening angles. 

With the values of $\theta_j$ known, it is then possible to correct the
isotropic equivalent $\gamma$-ray energy, $E_{\rm iso}(\gamma)$ for the effects
of conical geometry and derive the true $\gamma$-ray energy release 
(i.e., $E_\gamma = f_b \times E_{\rm iso}(\gamma)$ 
where 
$f_b \approx \theta_j^2 / 2 $). 
The
distributions of $E_{\rm iso}(\gamma)$ and $E_\gamma$ values are shown in 
Fig. 19. 
The somewhat surprising result is that $E_\gamma$ is tightly clustered around
$5 \times 10^{50}$ erg. 
Evidently, it appears that the central engines of GRBs
produce approximately a similar amount of energy, and a significant part, about
$10^{51}$ erg, escapes as $\gamma$-rays. 

\begin{figure}
\centerline{\psfig{figure=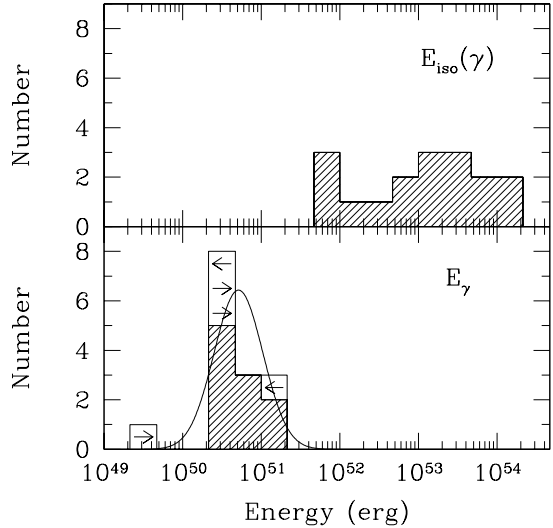,height=3.0in}}
\caption[]{
The distribution of the apparent isotropic $\gamma$-ray burst energy of GRBs
with known redshifts (top) versus the geometry-corrected energy for those GRBs
whose afterglows exhibit the signature of a non-isotropic outflow (bottom). 
The mean isotropic equivalent energy $\langle E_{iso}(\gamma)\rangle$ for 17
GRBs is $110 \times 10^{51}$ erg with a $1$-$\sigma$ spreading of a
multiplicative factor of 6.2.  In estimating the mean geometry-corrected energy
${\langle E_\gamma\rangle}$ we applied the Bayesian inference
formalism\cite{rei01a} and modified to handle datasets containing upper and
lower limits\cite{rlf+01}.  Arrows are plotted for five GRBs to indicate upper
or lower limits to the geometry-corrected energy.  The value of $\langle
\log{E_\gamma}\rangle$ is 50.71$\pm$0.10 (1$\sigma$) or equivalently, the mean
geometry-corrected energy $\langle E_\gamma\rangle$ for 15 GRBs is $0.5\times
10^{51}$ erg.  The standard deviation in $\log{E_\gamma}$ is
0.31$^{+0.09}_{-0.06}$, or a $1$-$\sigma$ spread corresponding to a
multiplicative factor of $2.0$.  From \cite{fr+01a}. 
}
\end{figure}

In addition to the implications that this result has on the luminosities and
energies of GRB central engines, it also affects determinations of the mean
beaming fraction and the GRB rate. Since conical fireballs are visible to only
a fraction, $f_b$, of observers, the true GRB rate, $R_t=\langle
f_b^{-1}\rangle R_{\rm obs}$, where $R_{\rm obs}$ is the observed GRB rate and
$\langle f_b^{-1}\rangle$ is the harmonic mean of the beaming fractions. We
find that the true GRB rate is $\langle f_b^{-1}\rangle\sim 500$ times larger
than the observed GRB rate, a result that provides some constraints for GRB
progenitor models.

\section{GRBs as Probes of Obscured Star Formation}

Already within months of the first detections of GRB afterglows, no OT's were
found associated with some well-localised bursts despite deep an rapid
searches; the prototype was GRB 970828.  There an XT was found, and a one-time
radio flare within its error circle (Fig. 20), which at the time was puzzling,
being different from the ``standard'' RT behaviour, as exemplified, e.g., by
GRB 970508.  But it was realised after the discovery of a radio flare from
GRB 990123 that such phenomena (ostensibly caused by reverse shocks) can indeed
point towards GRB afterglows.  GRB 970828 was thus the first case of a ``dark
burst'' (at least in the visible light) \cite{sgd+01b}.

\begin{figure}
\centerline{\psfig{figure=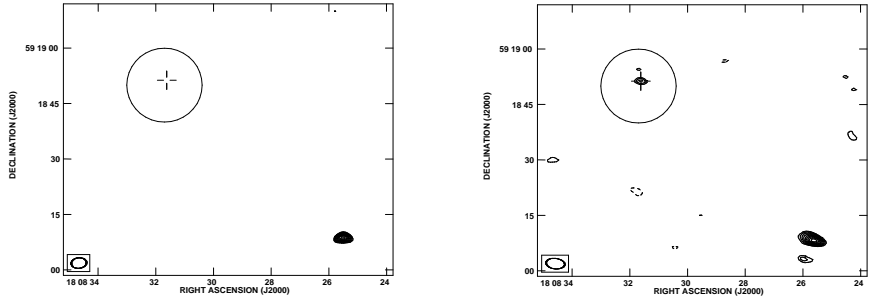,width=4.5in,angle=90}}
\caption[]{
The VLA maps showing the radio flare associated with GRB 970828, the
prototypical ``dark burst''.  The map on the left is from 31 Aug 1997 UT,
and the one on the right is from 01 Sep 1997 UT.  The circle indicates
the position of the X-ray afterglow.  The radio transient is positionally
coincident with a galaxy system shown in Fig. 21.
From \cite{sgd+01b}.
}
\end{figure}

Perhaps the most likely explanation for the non-detections of OT's when
sufficiently deep and prompt searches are made is that they are obscured by
dust in their host galaxies.  This is an obvious culprit if indeed GRBs are
associated with massive star formation.

Support for this idea also comes from detections of RTs without OTs, including
GRB 970828, 990506, and possibly also 981226 (see \cite{tbf+00,fbg+00}).
Dust reddening has been detected directly in some OTs 
(e.g., \cite{rkf+98,bfk+98,dkb+98} etc.);
however, this only covers OTs seen through optically thin dust, and there
must be others, hidden by optically thick dust.
An especially dramatic case was the RT \cite{tfk+98} and 
IR transient \cite{lg+98}
associated with GRB 980329.  We thus know that at least some GRB OTs must be
obscured by dust. 

\begin{figure}
\centerline{\psfig{figure=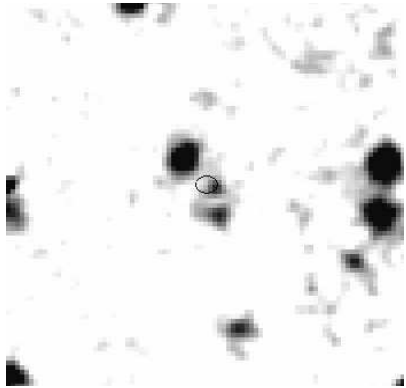,height=2.0in}}
\caption[]{
Keck $R$ band image of the host galaxy system of GRB 970828 at $z = 0.9579$.  
The morphology of the system is suggestive of a merger.
The ellipse indicates the position of the RT, close to what may be a dust lane
between two visible components.  Obscuration by dust is the most likely cause
of the non-detection of an OT associated with this burst.
From \cite{sgd+01b}.
}
\end{figure}

The census of OT detections for well-localised bursts can thus provide a
completely new and independent estimate of the mean obscured star formation
fraction in the universe.  Recall that GRBs are now detected out to $z \sim 4.5$
and that there is no correlation of the observed fluence with the redshift
(Fig. 12),
so that they are, at least to a first approximation, good probes of the star
formation over the observable universe.  As of mid-June 2001, there have been
$\sim 52 \pm 5$ adequately deep and rapid searches for OTs from well-localised
GRBs
\footnote{
We define ``adequate searches'' as reaching at least to $R \sim 20$ mag within
less than a day from the burst, and/or to at least to $R \sim 23 - 24$ mag
within 2 or 3 days; this is a purely heuristic, operational definition.  The
uncertainty comes from the subjective judgement of whether the searches really
did go as deep and as fast, based on what is published, mostly in GCN
Circulars, and whether the field was at a sufficiently low Galactic latitude to
cause concerns about the foreground extinction and confusion by Galactic stars.
}.
Out of those, $\sim 27 \pm 2$ OTs were found (the uncertainty being due to the
questionable nature of some candidates).
Some OTs may have been missed due to an intrinsically low flux, an unusually
rapid decline rate, or very high redshifts (so that the brightness in the
commonly used $BVR$ bands would be affected by the intergalactic absorption).
Thus the $maximum$ fraction of all OTs (and therefore massive star formation)
hidden by the dust is $(48 \pm 8)$\%.

This is a remarkable result.  It broadly agrees with the estimates that there
is roughly an equal amount of energy in the diffuse optical and FIR backgrounds
(see, e.g., \cite{mad99}).  This is contrary to some claims in the literature
which suggest that the fraction of the obscured star formation was much higher
at high redshifts.  Recall also that the fractions of the obscured and
unobscured star formation in the local universe are comparable.  GRBs can
therefore provide a valuable new constraint on the history of star formation
in the universe
\footnote{
These ideas were presented in talks at conferences and seminars by the members
of our group, starting in the spring of 1999.  They were then co-opted and
discussed in the same context by some other authors \cite{bn00}.
}.

There is one possible loophole in this argument: GRBs may be able to destroy
the dust in their immediate vicinity (up to $\sim 10$ pc?) \cite{wd00,gw00},
and if the rest of the optical path through their hosts ($\sim$ kpc scale?)
was dust-free, OTs would become visible.  Such a geometrical arrangement may
be unlikely in most cases, and our argument probably still applies.

Further support for the use of GRBs as probes of obscured star formation in
distant galaxies comes from radio and sub-mm detections of the hosts
\cite{fr+01b,bkf01}.

\section{Remaining Issues and Future Prospects}

It has been known for some years that GRBs exhibit a bimodal distribution
of durations\cite{kou+93}, 
which is also weakly correlated with the spectral hardness ratio (see Fig. 22).
Due to instrumental effects, $all$ of the GRBs with identified afterglows so
far are from the long/soft part of the distribution. 
It is likely that GRBs represent (at least) two distinct, although broadly
similar, physical phenomena.  The nature of the short/hard bursts, including
their progenitors, manifestations on other wavelengths, etc., remains completely
unknown as of this writing.  It is hoped that the forthcoming precise
localisations of such bursts by HETE-2 will help solve this problem.

\begin{figure}
\centerline{\psfig{figure=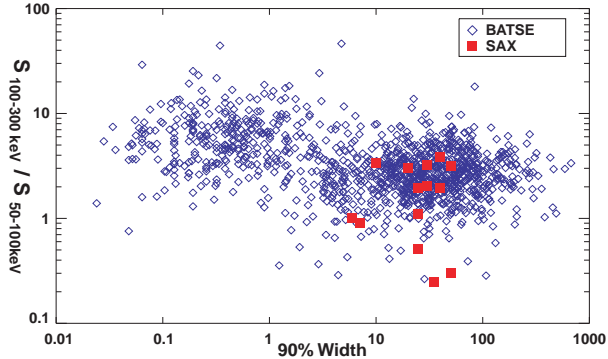,width=8.cm}}
\caption[]{
Distribution of duration ($T_{90}$) vs. spectral hardness for BATSE bursts
(diamonds) from the 4B catalogue \cite{pmp+97}. Events also localized by
BeppoSAX are shown by squares.  The bimodality of the distribution (which is
not an observational artifact) is evident.  Whether this corresponds to two
physically distinct phenomena or mechanisms is as yet uncertain.  It is also
unclear if there are more than two populations of GRBs.
}
\end{figure}

This begs the question of how many different types of bursts really are?  
Multivariate statistical studies of the BATSE data produced conflicting and
inconclusive results\cite{muk98,hak00}: there are at least two types (long
and short), but there may be more.  Moreover, if the identification of
GRB 980425 with SN1998bw is correct, that would represent yet another class
of sub-luminous GRBs (although it could also be interpreted as an orientation
effect).  There are also ``X-ray rich GRBs'', which may also be a distinct
population \cite{hei+01,kip+01}.

On the whole, we know very little about rapid variability (especially
transient, i.e., non-repeatable sources) at $any$ wavelength.  GRBs are
spectacular and exotic, drawing attention to themselves in the $\gamma$-ray
regime, but similarly interesting phenomena may exist with energy peaks at
lower frequencies.  In most surveys and observations (other than $\gamma$-ray
where photons are so scarce that anything counts) single-time events and
detections are routinely discarded (with some justification) as ``noise''.
We may be missing important new classes of rapid transient phenomena.
For example, serendipitous findings of fast, faint optical transients in
deep SN and lensing surveys\cite{schm+98,dlsweb,dier01}
may be examples of something new and interesting.  There is a great scientific
opportunity in dedicated, synoptic surveys of the sky at all wavelengths,
including (or especially?) optical.

Another exciting prospect is to use high-$z$ GRBs as cosmological probes. 
Even the present-day missions can in principle detect highly luminous (or
highly beamed) GRBs out to the expected epoch of initial galaxy formation and
reionization of the universe, at $z \sim 10 - 20$.  If GRBs are indeed 
associated with massive star formation, they may occur at such redshifts, and
be the most luminous objects in the universe then.  Straightforward
computations \cite{rl00}
show that their afterglows would be detectable by the existing technology. 
Searches in the radio regime may be especially promising \cite{cialo00}.
Such primordial GRBs would be predating the formation of luminous quasars
(whose massive central black holes may need a few $\times 10^8$ years to grow
and achieve comparable, albeit steady luminosities), and provide a unique
probe of the early IGM and the reionization epoch, as well as the early
star formation.  In this context, it is especially interesting that all modern
models of the primordial (i.e., metal-free) star formation suggest that the
IMF of the first stars would be dominated by very massive stars, i.e.,
precisely the now favorite progenitors of GRBs \cite{bcl99,lar99,abn00}.

While the studies of GRB afterglows and environments have generated much
progress in the physical understanding of the GRB phenomenon, and even
provided some strong hints about the possible progenitors, the nature of
the GRB ``central engines'' is still unknown.  In addition to the future
observations of gravitational waves with LIGO, LISA and other instruments,
two promising new information channel which may help us learn more about the
origin of GRBs are high-energy cosmic rays and neutrinos.  Both are 
expected in at least some models \cite{wax00,wb00}.
The probable detection \cite{atk+00} of TeV emission from GRB 970417a
may be the first glimpse of this new observational frontier.  Searches for
high-energy neutrinos with detectors like AMANDA
\footnote{ http://amanda.berkeley.edu/amanda/amanda.html}
are also very promising \cite{hh99}.

In this review we hoped to convey the richness and the excitement of the
field of GRBs, which now spans a broad range of astrophysical subjects.
In anything, the future of GRB studies seem even more exciting than the
spectacular progress of the past few years.  In addition to BeppoSAX 
\footnote{ http://www.asdc.asi.it/bepposax/}
and other current missions (e.g., 
Rossi XTE \footnote{ http://heasarc.gsfc.nasa.gov/0/docs/xte/},
HETE-2 \footnote{ http://space.mit.edu/HETE/},
Chandra \footnote{ http://chandra.harvard.edu/},
XMM-Newton \footnote{ http://xmm.vilspa.esa.es/},
and the IPN \footnote{ http://ssl.berkeley.edu/ipn3/}), 
future missions (e.g., 
SWIFT \footnote{ http://swift.sonoma.edu/},
GLAST \footnote{ http://glast.gsfc.nasa.gov/},
INTEGRAL \footnote{ http://astro.estec.esa.nl/SA-general/Projects/Integral/},
AGILE \footnote{ http://www.ifctr.mi.cnr.it/Agile/},
etc.)
as well as the ground-based networks of telescopes 
(e.g., REACT \footnote{ http://pulsar.ucolick.org/REACT/})
will continue to provide a steady stream of data and flashy discoveries in the
years to come.

\vspace*{-2pt}

\section*{Acknowledgments}
We wish to acknowledge the efforts of numerous collaborators worldwide, and the
expert help of the staff of Palomar, Keck, and VLA observatories and STScI. 
We also wish to emphasize critical contributions by the BeppoSAX team, whose
hard work and inspiration were essential in enabling the start of cracking of
the GRB puzzle.  This work was supported in part by grants from the NSF, NASA,
and private foundations to SRK, SGD, FAH, and RS, Fairchild Fellowships to RS
and TJG, Hubble Fellowship to DER, Millikan Fellowship to AD, and Hertz
Fellowship to JSB.  We also thank the organisers of the conference, especially
Drs. R. Ruffini and V. Gurzadyan for their support, hospitality, and patience.
This review combines contributions from several talks presented by the members
of our group.

\end{document}